\documentclass[fleqn,usenatbib]{mnras}

\usepackage{newtxtext,newtxmath}

\def\ltsima{$\; \buildrel < \over \sim\;$}
\def\ltsim{\lower.5ex\hbox{\ltsima}}
\def\gtsima{$\; \buildrel > \over\sim \;$}
\def\gtsim{\lower.5ex\hbox{\gtsima}}
\def\ms{$M_{\odot}$ }
\def\msp{$M_{\odot}$}

\usepackage[T1]{fontenc}

\DeclareRobustCommand{\VAN}[3]{#2}
\let\VANthebibliography\thebibliography
\def\thebibliography{\DeclareRobustCommand{\VAN}[3]{##3}\VANthebibliography}

\usepackage{graphicx}	
\usepackage{color}

\title[Core-Collapse Supernovae Limited to the Narrow Progenitor Mass Range]{From Galactic Chemical Evolution to Cosmic Supernova Rates Synchronized with Core-Collapse Supernovae Limited to the Narrow Progenitor Mass Range}

\author[T. Tsujimoto]{T. Tsujimoto$^{1}$\thanks{E-mail:
taku.tsujimoto@nao.ac.jp} 
\\
$^{1}$National Astronomical Observatory of Japan, Mitaka, Tokyo 181-8588, Japan
}

\date{Accepted XXX. Received YYY; in original form ZZZ}

\pubyear{2015}

\begin{document}
\label{firstpage}
\pagerange{\pageref{firstpage}--\pageref{lastpage}}
\maketitle

\begin{abstract}
Massive ($\geq8$\msp) stars perish via one of two fates:~core-collapse supernovae (CCSNe), which release synthesized heavy elements, or failed supernovae, thereby forming black holes. In the conventional Galactic chemical evolution (GCE) scheme, a substantial portion of massive stars, e.g., all stars in the mass range of 8$-$100 \msp, are assumed to enrich the Galaxy with their nucleosynthetic products. However, this hypothesis conflicts with the observations, namely, few CCSNe whose progenitor stars are more massive than $\sim$18 \msp. Here, we show that the chemical characteristics shaped by local thin disk stars are compatible with the predictions by enrichment via CCSNe limited to less massive progenitors in the new paradigm of Galactic dynamics that allows stars to migrate from the inner disk. This renewed GCE model predicts that the bursting star formation events$-$which are considered to take place in the Galactic bulge as well as in the thick disk$-$generate more numerous low-mass CCSNe than those expected from the locally determined canonical initial mass function. This finding suggests a high rate of CCSNe in early-type galaxies, which reflects a unique cosmic history of the CCSN rate. With considerable  contributions from these galaxies to the cosmic star formation rates in the early Universe, we predict a more steeply increasing slope of the CCSN rate with increasing redshift than that in proportion to cosmic star formation. This predicted redshift evolution agrees well with the measured rates for 0 \ltsim $z$\ltsim 0.8; however, its predicted CCSN rate for higher-$z$ calls for more precise data from future surveys. 
\end{abstract}



\section{Introduction}

Core-collapse supernovae (CCSNe) represent the final phenomenon exhibited by massive ($>$ 8\msp) stars and play a major role in the  chemical enrichment in galaxies. However, not all massive stars end with the formation of CCSNe;~indeed, a small fraction of these stars  fail to explode and instead form black holes (BHs), which do not contribute to the chemical evolution of the surrounding galaxy \citep{Kochanek_08}. In general, however, the existing Galactic chemical evolution (GCE) models do not seriously consider the fraction of stars that becomes BHs, $f_{\rm BH}$;~rather, they assume that chemical enrichment proceeds by all massive stars with masses typically reaching up to 50\ms \citep[e.g.,][Kobayashi et al.~2006\footnote{In \citet{Kobayashi_20}, the concept of $f_{\rm BH}$ is introduced into the models, but the CCSNe of stars with masses up to 50\ms are allowed as hypernovae.}]{Tsujimoto_95}, 100\ms \citep[e.g.,][]{Masi_18}, or even 120\ms \citep[e.g.,][]{Prantzos_18}. Though some earlier works \citep[e.g.,][]{Raiteri_91, Portinari_98} already design the GCE models with $f_{\rm BH}$, considering the stars collapsing to BHs with their masses such as 40\msp, these theoretical approaches so far recently confront growing alienation from the observations: Strong evidence from the identification of presupernova stars indicates that the most massive CCSN progenitors have masses of only approximately 18\msp; that is, the masses of CCSN progenitors should be narrowly confined to 8$-$18\ms \citep{Smartt_09, Smartt_15}. The hypothesis that more massive stars will explode as stripped SNe \citep[e.g.,][]{Ekstrom_12} as the result of increased mass-loss from existing red supergiants with the mass of $\sim$25\ms seems implausible since higher mass-loss rates than observed are required \citep[e.g.,][]{Beasor_20}. In addition, such a low implied upper bound for the mass of a CCSN progenitor is consistent with the theoretical modeling of CCSNe in terms of their explodability \citep{Ugliano_12, Sukhbold_16, Kresse_21}, which is correlated with the compactness of stellar cores \citep{Connor_11}. This case naturally results in a high $f_{\rm BH}$ ($\sim$30\%) with a possible range of 4$-$39\% constrained by estimates based on a search for failed supernovae \citep{Neustadt_21}. 

Therefore, it might be attractive to establish a theoretical framework in which the upper mass bound of CCSN progenitors ($m_{\rm cc, u}$) is no longer a free parameter and is instead fixed at approximately $m_{\rm cc, u}$= 18 \msp. This fixed $m_{\rm cc, u}$ precisely gives the number of CCSNe in each generation of stars as long as the locus is specified to the solar vicinity. This is because the birth mass distribution of massive stars between the lower mass bound ($m_{\rm cc, l}$ = 8\msp) and $m_{\rm cc, u}$, referred to as an initial mass function (IMF), is well approximated by a power law with a slope index, denoted $x$ (=-1.35:~the Salpeter)\footnote{The Kroupa's IMF gives  the almost same power index of $x$=-1.3 for $m$>1 \ms \citep{Kroupa_01}.}. The impact on the GCE caused by a reduction in $m_{\rm cc, u}$ from $\sim$100\ms to 18\ms is sufficiently large. This is due not only to the fact that the CCSN frequency reduces to $\sim$ 70\%, but also to the fact that more heavy elements are generally ejected from CCSNe whose progenitor stars are more massive with a larger core mass \citep[e.g.,][]{Woosley_95}.
 
Then, the question arises as to whether the predictions of GCE models with $m_{\rm cc, u}$= 18\ms match the observed chemical abundances exhibited by nearby stars. The difficulty of addressing this issue is alleviated by the renewed views regarding the chemodynamical evolution of the Galaxy. The improved understanding of Galactic dynamics suggests that stars radially move on the disk when they encounter transient spiral arms that are naturally generated during the process of disk formation \citep[e.g.,][]{Sellwood_02, Roskar_08}. This so-called radial migration of stars predicts that the stars in the solar vicinity represent the mixture of stars born at various Galactocentric distances over the disk. In particular, this dynamic process induces a major migration from the inner disk, which forms faster and becomes more metal-rich than the solar vicinity according to the inside-out scenario \citep{Chiappini_01}. Thus, it is vital to update GCE models to consider that stars born under less efficient CCSN enrichment than previously thought owing to a low $m_{\rm cc, u}$ contribute to only a part of the local Galactic chemistry and that the remaining composition must be due to more efficient enrichment trajectories than an in situ one. 

Then, given the confirmed validity of GCE models with a low $m_{\rm cc, u}$, it is natural to extend our attention beyond the local field  to the Galactic bulge, which is a generally metal-rich population that forms within a short timescale of \ltsim 2 Gyr \citep{Barbuy_18}. Such Galactic bulge's properties suggest that CCSNe play a more vital role in the enrichment that rapidly proceeds in the bulge compared to the slow progress in the Galactic disk since, for many type Ia supernovae (SNe Ia) whose delay times are on the order of billions of years,  the star ceases forming before the release of heavy elements. This consideration could support our expectation that the number of CCSNe per stellar generation deduced from the local IMF having $x$=-1.35 with $m_{\rm cc, u}$= 18 \ms is not sufficient to reproduce the chemical characteristics of the Galactic bulge. This argument would be plausible since the existing GCE models (even those with $m_{\rm cc, u}$= 100 \msp) suggests a flatter IMF to better fit the observations of the bulge \citep[e.g.,][but see Bensby et al.~2017]{Matteucci_90, Ballero_07}.

Knowledge of the IMF in the Galactic bulge could be transferred to the study of other spheroids, i.e., the bulges of other spiral galaxies and elliptical galaxies in terms of similar chemistries \citep[e.g.,][]{Thomas_06} as the result of a common chemical evolutionary history, which can be summarized as fast chemical enrichment leading to a mean metallicity up to (or exceeding) the solar metallicity within a few billion years at most. We stress that the possibility of variation in the IMF significantly increases by a potentially low $m_{\rm cc, u}$ under an ongoing intense debate on the universality versus nonuniversality of the IMF \citep{Bastian_10, Hopkins_18}.

The controversy regarding whether the IMF varies among different types of galaxies can be explored by comparing the occurrence rates of  star formation and CCSNe in the Universe as a function of redshift ($z$), that is, whether the measured rate of CCSNe ($R_{\rm cc}$) is proportional to the cosmic star formation rate (denoted $\psi$). Since the contributing fraction in $\psi$ from individual types of galaxies varies in accordance with $z$, the IMF variation, if it exists, should lead to a break in the proportionality relating $\psi$ to $R_{\rm cc}$. In fact, the measured $R_{\rm cc}$$-$$z$ trend detaches from a $\psi$$-$$z$ relation;~the contrast of $\psi$ between the present and $z\approx 1$ is below a factor of 10 \citep{Hopkins_06, Madau_14, Davies_16, Driver_18}, whereas $R_{\rm cc}$ for $z>$ 0.5 \citep[e.g.,][]{Petrushevska_16} suggests a higher rate at $z\approx1$ than the current one by more than a factor of 10. In this paper, based on a low $m_{\rm cc, u}$ motivated both observationally and theoretically, we first discuss GCE and then explore the CCSN rate history of the Universe.

\section{Galactic chemical evolution} 

First, we examine the chemical evolution of the Galactic disk composed of two chemically distinguishable populations, namely, thin and thick disks, and a bulge;~for this purpose, we adopt an IMF with ($m_{\rm cc, l}$, $m_{\rm cc, u}$) = (8 \msp, 18 \msp). In this scheme, massive stars ($m>$ 18 \msp) are assumed not to contribute to Galactic chemical enrichment due to failed supernovae and the formation of  BHs. We calculate the evolution of two elements: Fe and Mg. For the Fe yield from CCSNe, we adopt 0.06 \ms based on recent observational estimates from luminosities \citep{Rodriguez_21}. Then, we deduce a Mg yield of 0.08 \ms from an observed plateau of [Mg/Fe] $\approx+0.4$ among halo stars, which reflects the average nucleosynthesis Mg/Fe ratio among CCSNe;~this value is within theoretical predictions \citep{Tominaga_07}. The star formation history (SFH) of each Galactic component is modeled by changing the timescale of star formation ($\tau_{\rm SF}$) and the supply of gas from the halo ($\tau_{\rm in}$) for a given duration $\Delta_{\rm SF}$. Detailed description of GCE models are given in the following section.

\subsection {GCE Models}

\noindent$-$the thin disk$-$

The basic picture is that the thin disk was formed through a continuous low-metal infall of material from outside the disk region (i.e., the inter-galactic medium) based on the inside-out formation scenario \citep{Chiappini_01}, that is, the disk is formed by an infall of gas occurring at a faster rate in the inner region than in the outer ones consistent with the shorter dynamical times. Here, we calculate chemical evolution at three regions with their Galactocentric distances $R_{\rm GC}\approx$ 4 (inner disk), 8 (solar vicinity), and 12 kpc (outer disk). This assignment of $R_{\rm GC}$ for each model can be done by comparing the predicted achieving [Fe/H] values with the current [Fe/H]-$R_{\rm GC}$ relation (i.e., the observed radial [Fe/H] gradient) \citep{Genovali_14}, which approximately gives [Fe/H]$\approx$0.2-0.3 at $R_{\rm GC}\approx$ 4 kpc and [Fe/H]$\approx$-0.15 at $R_{\rm GC}\approx$ 12 kpc. We calculate the gas fraction and the abundance of heavy-element in the gas at each region with an assumed IMF, $\phi(m)$, for a mass range from $m_l$ = 0.01 \ms to $m_u$ = 100 \msp.

\begin{figure*}
	\vspace{0.5cm}
	\hspace{-0.7cm}
	\includegraphics[width=1.78\columnwidth]{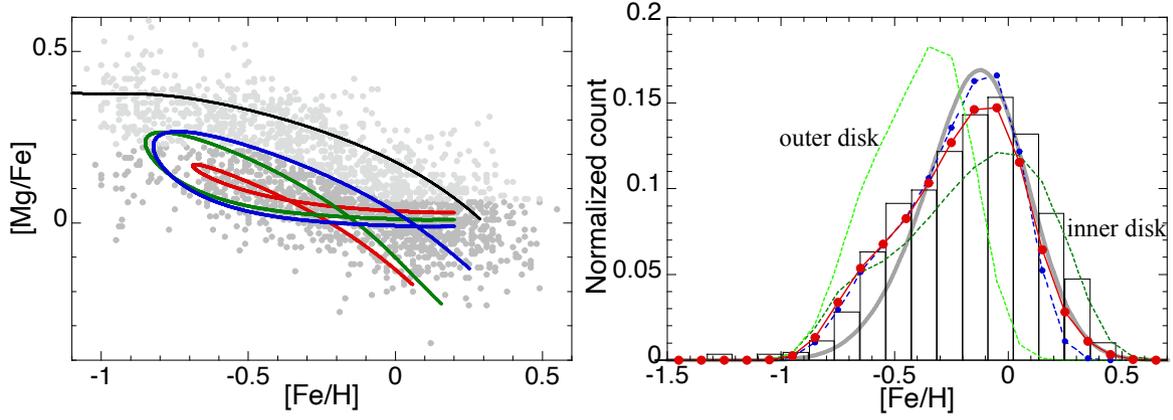}
    \caption{Left: Calculated correlation of [Mg/Fe] with [Fe/H] for disk stars in the solar vicinity compared with observed data from the Stellar Abundances for Galactic Archeology (SAGA) database. The data are separated into thick (light gray) and thin (dark gray) disk points using the boundary between the two defined by \citet{Hayden_17};~however, this boundary is slightly modified according to the result of \citet{Schultheis_17}. The four colored curves are the model results for the thin disk (red: in situ, green: the inner disk, blue: the inner disk with $x$=-0.9: light  green: the outer disk), while the black curve is for the thick disk with $x$=-0.9. The initial [Mg/Fe] of gas for the thin disk is set with an offset of 0.02 dex among the four. Right: Comparison of the predicted local MDFs with the observed MDFs (gray curve: Nandakumar et al.~2017; histogram: Thompson et al. 2018). The predicted MDFs correspond to the in situ (dashed blue curve),  inner disk (dotted green curve), and outer disk (dotted light green curve) functions, and the mixture of these three MDFs with a ratio of 6.5:2.5:1 is plotted as the red curve.}
\end{figure*}

Let $\psi(t)$ be the star formation rate and $A(t)$ be the gas infall rate, then the gas fraction $f_g(t)$ and the abundance of heavy-element $i$ $Z_i(t)$ in the gas at each region change with time according to 
\begin{equation}
\frac{df_g}{dt}=-\psi(t)+\int^{m_{u}}_{{\rm max}(m_l,m_t)}dm\phi(m)r(m)\psi(t-t_m)+A(t)
\end{equation}
\begin{eqnarray}
\hspace{-2cm}\frac{d(Z_if_g)}{dt}=-Z_i(t)\psi(t)+\int^{m_u}_{{\rm max}(m_l,m_t)}dmA\psi(m)y_{{\rm Ia},i} \nonumber \\
\times \int^t_0 dt_{\rm Ia}g(t_{\rm Ia})\psi(t-t_{\rm Ia}) \nonumber \\
+\int^{m_{u}}_{{\rm max}(m_l,m_t)}dm (1-A)\phi(m)[y_{{\rm cc},i}+Z_i(t-t_m) \nonumber \\
r_w(m)]\psi(t-t_m) +Z_{A,i}(t)A(t) \ \ ,
\end{eqnarray}
\noindent where $m_t$ is the turnoff mass when the main-sequence lifetime, $t_m$, is equal to time $t$, $r(m)$ is the fraction of the ejected material from a star of mass $m$, $r_w(m)$ is the fraction of the ejected material without newly synthesized elements from that star, $y_i$ is the heavy-element yield from an CCSN or SN Ia, and $Z_{A,i}$ is the abundance of heavy element contained in the infalling gas.

The star formation rate, $\psi(t)$, is assumed to be proportional to the gas fraction with a constant coefficient of $1/\tau_{\rm SF}(R)$, where $\tau_{\rm SF}(R)$ is a timescale of star formation as a function of $R$. For the infall rate $A(t)$, we adopt the formula that is proportional to $t\exp(-t/t_{\rm in}(R))$ with a timescale of infall of $\tau_{\rm in}(R)$. According to the inside-out scenario, $\tau_{\rm SF}(R)$ and $\tau_{\rm in}(R)$ are assumed to increase outwards, and the adopted timescales together with the duration of star formation $\Delta_{\rm SF}(R)$ in unit of Gyr are ($\tau_{\rm SF}$, $\tau_{\rm in}$, $\Delta_{\rm SF}$)=(0.5, 1, 3.5), (1, 5, 10), and (3.3, 10, 10), respectively. The metallicity, $Z_{A,i}$, of an infall is assumed to be very low-metallicity ([Fe/H]=-1.5), which is implied by the metallicity measurement of damped Ly$\alpha$ systems \citep{Wolfe_05} with a SN-II like enhanced [Mg/Fe] ratio (=0.4). 

As the initial abundances, we assume the high value of [Fe/H] (=0.2) together with [Mg/Fe]$\sim$0 including some variation as the thick disk's remaining gas \citep[see also][]{Spitoni_19}. Here, we regard the thick disk as the first disk, which is heated up by an ancient merger such as Gaia-Enceladus \citep{Helmi_18}, that is subsequently followed by the gradual formation of a secondary disk, i.e., the thin disk. Such a first thick disk can also be formed through clump merging in an unstable primordial disk \citep{Bournaud_07}. In these scenarios, star formation within the thin disk could occur after the termination of star formation in the thick disk.

The adopted nucleosynthesis yields of CCSNe for Fe and Mg are deduced from the observational bases:~the SN light curve and the plateau of abundance ratio for halo stars, as already stated. These nucleosynthesis products are released with a short delay time corresponding to the lifetimes of massive stars. For SNe Ia, each event is assumed to ejects 0.63 \ms of Fe and 8.5$\times 10^{-3}$ \ms of Mg \citep{Iwamoto_99}   
according to the delay time distribution (DTD), $g(t_{\rm Ia})$, which is proportional to $t_{\rm Ia}^{-1}$ with a range of 0.1 $\leq$ $t_{\rm Ia}$ $\leq$ 10 Gyr \citep{Maoz_14}. The DTD is normalized so that 8\% of the primary stars in binaries with initial masses in the range of 3-8 \ms explode as SNe Ia: $A$=0.08 for 3-8 \ms and $A$=0 outside this mass range. Its fraction has been obtained through previous works \citep[e.g.,][]{Tsujimoto_12} and has been rechecked by this study.

\noindent$-$the thick disk and the bulge$-$

Chemical evolutions of the thick disk and the bulge are calculated in the scheme of a relatively rapid star formation by adopting a short timescale of star formation with a rapid collapse. These are parameterized by ($\tau_{\rm SF}$, $\tau_{\rm in}$, $\Delta_{\rm SF}$)=(0.7, 1, 2.5) and (0.25, 0.5, 2.5), respectively. The two Galactic components start to be formed from a low-metallicity infalling gas without an  initial gas. The adopted short $\tau_{\rm SF}$ is within the predicted values by previous studies: 0.5-1 Gyr \citep{Kobayashi_06} for the thick disk and 0.05-0.5 Gyr \citep{Matteucci_90, Kobayashi_06, Ballero_07, Grieco_12} for the bulge. These $\tau_{\rm SF}$ values adopted in the models can be compared with the observed molecular depletion time in galaxies. In the active star-formation phase for 1 \ltsim $z$ \ltsim 5, i.e., $\sim$8-13 Gyr ago, $\tau_{\rm SF}$ of galaxies in the PHIBSS survey is strictly confined less than 1 Gyr \citep{Tacconi_18} with their median values reanalyzed by \citet{Segovia_22} being approximately 0.4-0.5 Gyr, and thus broadly agrees with our adopted values. Here, we note that the cosmological simulations predict that $\tau_{\rm SF}$ mainly settles down to $\approx$0.2-0.3 Gyr during a starburst phase for $z$\gtsim 1 in Milky Way-like galaxies \citep{Segovia_22}. On the other hand, $\tau_{\rm SF}$ at a secular stage of star formation increases to $\approx$1-2 Gyr \citep{Leroy_13, Tacconi_18} as measured in nearby galaxies; this timescale is comparable to the values at the solar vicinity (=1 Gyr) or at the outer disk (=3.3 Gyr) in our model.

\subsection{Galactic disk}

\begin{figure}
	\vspace{0.3cm}
	\includegraphics[width=0.9\columnwidth]{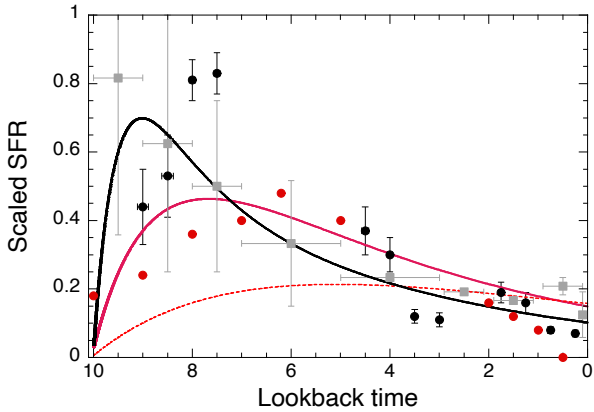}
    \caption{Predicted and observed SFHs in the solar vicinity. The modeled SFH (including the migrators from the inner and outer disks) is indicated by the black curve and is compared with observed data \citep[black circles:][gray squares: Sahlholdt et al. 2022]{Alzate_21}, the in situ SFHs obtained by the modeling (high-SFR model:~solid red curve;~low SFR model:~dashed red curve), and the corresponding observations \citep[red circles:][]{Sahlholdt_22}.}
\end{figure}

Here, we focus mainly on comparing theoretical predictions with the observations of thin disk stars that are currently present in the solar vicinity. For these stars, we adopt the Salpeter IMF ($x$=-1.35) for the entire mass range of 0.01$-$100 \msp. In the modeling, the following three aspects should be highlighted:~(i) a higher star formation rate (SFR) in the solar vicinity than that previously adopted should be assigned to supplement inefficient enrichment by CCSNe owing to a low $m_{\rm cc, u}$;~(ii) the chemical evolution of the thin disk likely started from a high metallicity since an end product in the formation of a thick-disk is a metal-rich gas; and (iii) more efficient enrichment paths (leading to a faster increase in metallicity and higher metallicity than that in the solar vicinity) should be added to some local chemical characteristics to include the effect of migrators from the inner disk.

\begin{figure*}
	\vspace{0.5cm}
	\hspace{-0.7cm}
	\includegraphics[width=1.78\columnwidth]{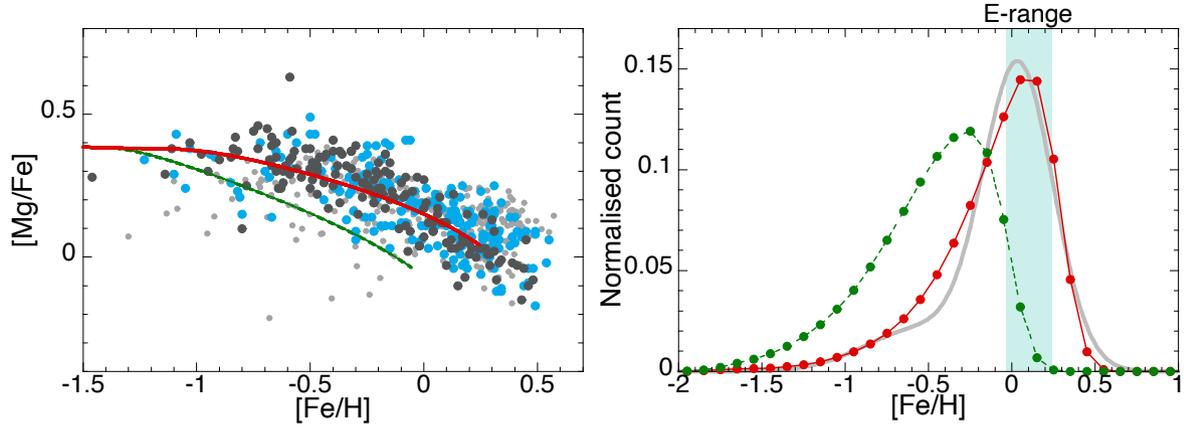}
    \caption{Chemical evolution of the Galactic bulge. We calculate two cases with different IMF slopes: $x$=-0.9 (red curve) and $x$=-1.35 (dashed green curve). Left:~Calculated correlation of [Mg/Fe] with [Fe/H], compared with the observed data \citep[black circles: Gonzalez et al. 2011; cyan circles:][gray dots: Schultheis et al. 2017]{Johnson_14}. Right: Comparison of the predicted MDFs with the observed MDF \citep[gray curve:][]{Schultheis_21}, which closely resembles the result from the Blanco DECam Bulge Survey \citep{Johnson_20}. The observed average [Fe/H] range exhibited by elliptical galaxies whose stellar masses are larger than $10^{10}$\ms \citep{Arrigoni_10} after  being converted from [Z/H] and [$\alpha$/Fe] is indicated by the light cyan stripe.}
\end{figure*}

Incorporating (i) and (ii), we model the one for the solar vicinity by ($\tau_{\rm SF}$, $\tau_{\rm in}$, $\Delta_{\rm SF}$)=(1, 5, 10) Gyr in conjunction with initial abundances of [Fe/H]=+0.2 and [Mg/Fe]=+0.02. Note that $\tau_{\rm SF}$=3.3 Gyr is assigned for the case with $m_{\rm cc, u}$=50\ms \citep{Tsujimoto_21}. The calculated evolution of [Mg/Fe] is shown by the red curve in the left panel of Figure 1. We see that [Fe/H] and [Mg/Fe] first decrease and increase, respectively, owing to dilution by metal-poor infalling gas;~subsequently,  this reverse evolution comes to an end when the chemical enrichment by star formation exceeds the effect of gas dilution, after which a typical evolutionary path appears. While this single trajectory encompasses only a relatively small part of the data (dark gray points), an end result of (iii)$-$i.e., this outcome in addition to an additional track (green curve) predicted for the inner disk ($R_{\rm GC}\approx$ 4kpc) representing fast enrichment by ($\tau_{\rm SF}$, $\tau_{\rm in}$, $\Delta_{\rm SF}$)=(0.5, 1, 3.5) Gyr$-$occupies a wider data-populated area, yielding a more consistent view regarding the validation of our GCE models. In addition to such contribution from the inner disk stars to the local chemistry, we also draw the evolutionary track, which is expected for the migrators from the outer disk around $R_{\rm GC}\approx$ 12kpc (light green curve) since numerical models predict inward migration from the outer disk as well, though its  efficiency is much lower than that from the inner disk \citep[e.g.,][]{Roskar_08}. In these models, we do not consider the loss of metals by the lock-up in the hot interstellar medium as well as an outflow to the inter-galactic medium. We confirm that its loss rate of about 10-15\% still gives agreement with the observations \citep[see][]{Schonrich_19}.

Moreover, we consider another inner disk model invoking more numerous CCSNe by adopting a flatter IMF ($x$=-0.9), which is also in good agreement with the observations (blue curve). In this model, the chemical evolution of the thick disk is reproduced via rapid formation by the same flat IMF modeled by ($\tau_{\rm SF}$, $\tau_{\rm in}$, $\Delta_{\rm SF}$)=(0.7, 1, 2.5) Gyr (black curve). Thanks to the  similarity of the [Mg/Fe]-[Fe/H] correlation between the thick disk and the bulge \citep{Alves_10}, we leave the argument regarding its chemical evolution to that of the bulge in the next section. Here, we note simply that the [Mg/Fe] evolution of the thick disk demands a flat IMF.

Paralleling the evolution of the abundance ratio, the observed metallicity distribution function (MDF) imposes stringent constraints on the models. The right panel of Figure 1 shows a comparison between the predicted MDFs (colored curves) and the observed MDF (gray curve and histogram), demonstrating that the MDF built by an in situ star formation (dashed blue curve) alone is approximately in agreement with the observations. The hybrid model (solid red curve) composed of three MDFs, one in situ and ones in the inner disk (dashed green curve) and in the outer disk (dashed light green curve), with a mix ratio of 6.5:2.5:1 \citep[cf.,][]{Roskar_08} matches the observed distribution as well. Here, it should be noted that our simple assessment of GCE based on one-zone chemical evolution models should be extensively explored by more sophisticated models including the detailed effects from radial migration \citep[e.g.,][]{Schonrich_09, Kubryk_15, Frankel_18}.

Finally, we remark on the SFH. As already discussed, a GCE model with a low $m_{\rm cc, u}$ demands a relatively high SFR, which predicts enhanced star formation at the early phase of disk formation. The resultant features together with the observations are shown in Figure 2, which plots two results:~an in situ SFH (red curve) and a synthesized SFH including the effect of migration in the same manner as for the MDF (black curve), both of which are approximately compatible with the observed trends corresponding to the same colors. It should be stressed that a low-SFR model ($\tau_{\rm SF}$=3.3 Gyr), which is designed for the case with a high $m_{\rm cc, u}$ (e.g., 50 \msp), predicts a nearly flat evolution (dashed red curve), indicating its large deviation from the observed SFH.

\subsection{Galactic bulge}

According to the compatibility of GCE models featuring a low $m_{\rm cc, u}$ (=18\msp) with the chemical characteristics observed from nearby disk stars, we next apply our model to the Galactic bulge. As anticipated by the results for the thick disk, the observed correlation of [Mg/Fe] with [Fe/H] in the bulge demands a flatter IMF ($x$=-0.9) from the model, as shown by the red curve in the left panel of Figure 3. Here, we adopt ($\tau_{\rm SF}$, $\tau_{\rm in}$, $\Delta_{\rm SF}$)=(0.25, 0.5, 2.5) Gyr. On the other hand, the modeling result with the Salpeter IMF (dashed green curve) predicts an excessively early break in [Mg/Fe] at [Fe/H] $\ll$-1 and detaches from the main stream supported by data. Such a serious discrepancy between this model and the observations (the gray curve) is also seen in the MDF;~at the same time, however, an excellent match for the flat IMF model is confirmed (the right panel). If we assume a larger value of $\tau_{\rm SF}$ such as 0.5 Gyr, which is more comparable to the observation \citep{Tacconi_18}, our model needs a slightly more flat IMF, i.e., $x>-0.9$, to compensate a less heavy-element production due to a slower star formation.

Furthermore, it is worth while to discuss the deduced amount of IMF variability in the theoretical framework of the complex explosion/black hole landscape \citep{Sukhbold_16}. Supernova explosion models do not predict a clear-cut $m_{{\rm cc},u}$ since the explodability of massive stars is not a simple function of stellar mass. For instance, one of their models (named N20) exhibits a explodability mapping showing a major $m_{{\rm cc},u}$ around 22 \ms with a lack of $\sim$15 \ms stars and an island of explosion for the stars with $\sim$ 25.5-27.5 \msp. Another model (named W18) exhibits a feature similar to N20 but with an additional lack of explosion for the stars with $\sim$19 \msp. For these two cases, we estimate the degree of change in the IMF slope from the case with $m_{{\rm cc},u}$ = 18 \msp. The two explodability mappings increase the number of CCSNe by 13\% and 11\%, each of which is equivalent to $m_{{\rm cc},u}$ = 22.6 \ms and 21.2 \msp, respectively, if we assume a single mass range starting from $m_{{\rm cc},l}$ = 8\msp. Then, such small increases in the CCSN rate per each generation of stars by the two cases result in the requited IMF slope of $x$=-0.96, -0.95, respectively, in the bulge, that corresponds to a slight change in the slope by $\sim$0.05. Accordingly, we conclude that a flat IMF in the bulge is guaranteed even under the condition given by the complex explodability mapping. 

The argument for a flat IMF in the Galactic bulge can be extended to an insight into the form of the IMFs in elliptical galaxies (Es). A mean metallicity among Es (0\ltsim$\langle$[Fe/H]$\rangle$\ltsim0.2:~cyan stripe) resulting from a bursting SFH in common with the bulge indicates a flat (i.e., top-heavy) IMF in Es \citep[e.g.,][]{Arimoto_87}. Then, if we incorporate the results suggesting a flat IMF for the thick disk (and possibly for the innermost disk in a similar manner), we naturally conclude that an IMF generating numerous massive stars could emerge from the bursting star formation events \citep{Pouteau_22}. 

\section{Cosmic history of core-collapse supernova rates}

Based on the understanding of chemical evolution of the different components of the Galaxy, we next explore the redshift evolution of the CCSN rate ($R_{\rm cc}$) traced by the cosmic SFR ($\psi$). We calculate $R_{\rm cc}$ by converting it from the observationally estimated $\psi$ via a scale factor, $k_{\rm cc}$, of massive stars that explode as CCSNe per unit mass of the IMF, which is related by 
\begin{equation}
\hspace{2cm}R_{\rm cc} ($z$) = k_{\rm cc}(z) \ h^2 \ \psi(z) \ \ . 
\end{equation}

\noindent Here, $h$ is the Hubble parameter, and the units of $R_{\rm cc}$ and $\psi$ are yr$^{-1}$ Mpc$^{-3}$ and \ms yr$^{-1}$ Mpc$^{-3}$, respectively. In our calculations, $k_{\rm cc}$ varies in accordance with $z$ since the $\psi(z)$ value is a composite of the SFRs contributed from different types of galaxies in which the SFH differs and the IMF is assumed to vary. Thus, each type $j$ of galaxies has a unique value of $k_{\rm cc}$, i.e., $k_{{\rm cc},j}$.


To evaluate $k_{\rm cc} (z)$, we classify galaxies into five groups;~spheroids (E/S0) and four classes of spiral galaxies (Sab, Sbc, Scd, and Sdm). Then, given their SFHs, the relative contribution to $\psi (z)$ from each type of galaxy is calculated by weighting with its relative proportion and mass-to-luminosity ratio \citep{Totani_96}. Then, $k_{\rm cc}(z)$ is defined as

\begin{equation}
\hspace{2cm}k_{\rm cc}(z)=\sum_{j=1}^5 k_{{\rm cc},j}\frac{w_j \psi_j(z)}{\sum_{j=1}^5 w_j \psi_j(z)} \ \ ,
\end{equation}

\noindent where $w_j$ and $\psi_j(z)$ are the individual weight and the SFR for the $j$-type galaxy, respectively. Here, we set the formation redshift of galaxies at $z$=2.5 to match the observational trend of $\psi$ for $z<$ 2:~a gradual decrease from its peak at $z\approx2$ to the present \citep[e.g.,][]{Madau_14}. For the cosmological parameters, we adopt $\Omega_\Lambda$=0.69, $\Omega_M$=0.31, and $H_0$=67.74 km s$^{-1}$Mpc$^{-1}$.

In this study, we adopt the Salpeter IMF for late-type galaxies (Sbc, Scd, and Sdm) and the flat IMF ($x$=-0.9) for early-type of galaxies (E/S0 and Sab) to count the number of CCSNe whose progenitor masses are in the range of 8-18\msp. The assignment of a flat IMF to Sab is attributable to our findings regarding the Galactic disk components corresponding to the thick disk (and possibly the innermost thin disk)\footnote{To ensure that Sab-type galaxies are better suited to a flatter IMF, relatively rapid formation on a 2-Gyr timescale of star formation is assumed, which is modified from the SFH adopted by \citet{Totani_96}.}. 
Given the entire mass range of 0.01$-$100\ms with the single slope of the Salpeter IMF, $k_{\rm cc}$ is deduced to be $\sim 0.05$ $M_\odot^{-1}$. Considering the  uncertainty in the complex mass distribution for low-mass stars ($m<$1 \msp) \citep{Kroupa_01, Chabrier_03}, we adopt a slightly modified $k_{\rm cc}$ of 0.06 $M_\odot^{-1}$ for late-type galaxies, which yields $k_{\rm cc}$ = 0.18 $M_\odot^{-1}$ for a flat IMF ($x$=-0.9) for  early-type galaxies.

\begin{figure}
	\vspace{0.3cm}
	\includegraphics[width=0.9\columnwidth]{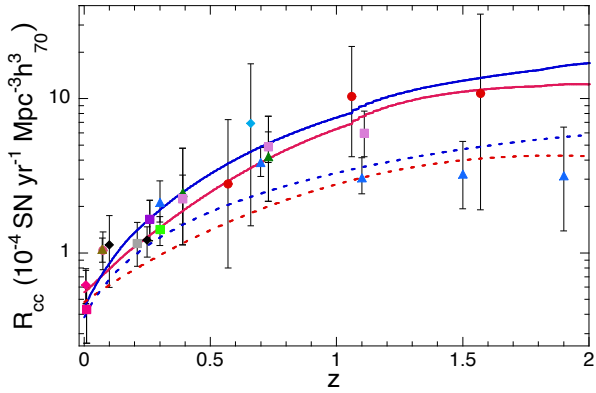}
    \caption{Cosmic CCSN rates predicted with respect to the redshift for $z<2$ and compared with the measured rates. The theoretical curves are calculated based on the cosmic SFRs deduced from \citet[][red curves]{Madau_14} and \citet[][blue curves]{Hopkins_06}. The models corresponding to the solid curves hypothesize a large number of CCSNe in early-type galaxies, while an equal number of CCSNe against the total stellar mass among all types of galaxies is assumed for the dotted curves under a conventional scheme. The measured rates for the high-redshift cases ($z>1$) are from \citet[][red circles]{Petrushevska_16} and \citet[][blue triangles]{Strolger_15}. For $z$\ltsim1, the data points are assembled from various sources \citep{Cappellaro_15}.}
\end{figure}

Thus, we finally deduce $R_{\rm cc} (z)$ by combining $k_{\rm cc} (z)$ with the measured $\psi (z)$. The results corresponding to $z\leq2$ based on two choices of $\psi$~\citep{Madau_14, Hopkins_06} are shown in Figure 4 as the solid red and blue curves, respectively. For comparison, the cases with a constant $k_{\rm cc}$ (= 0.06 $M_\odot^{-1}$), which is supported under a universal IMF, are indicated by dashed curves. Clearly, the observed $R_{\rm cc}$ trend for $z$\ltsim 1 is shown to be in good agreement with the modeling results assuming  a varying $k_{\rm cc}$, whereas the models with a constant $k_{\rm cc}$ fail to reproduce well a contrast of $R_{\rm cc}$ between the present and $z\approx0.8$. The enlarged contrast obtained by adopting a varying $k_{\rm cc}$ results from a switch of the dominant galaxy population contributing to $\psi$ from a high-$z$ Universe to a low-$z$ Universe, that is, E/S0 at $z\approx2$ (nearly 100\%) to late-type galaxies including our own at $z\approx0$ ($\approx$94\%). However, the results for $z>$1 by the models with a varying $k_{\rm cc}$ are in noticeable tension with the measured rate by \citet{Strolger_15}. Since another high-$z$ data by \citet{Petrushevska_16} supporting our model involve the substantial uncertainties, the overall trend for 0$<$$z$$<$2 shaped by existing data does not rule out the possibility of a constant $k_{\rm cc}$ model, preserving the argument by \citet{Strolger_15} that rejects a redshift-evolving IMF \citep{Dave_08} as well as the $k_{\rm cc}$ value derived from $m_{\rm cc, u}<$ 20 \msp. In addition, clear answers as to whether of a variable or constant $k_{\rm cc}$ elude us also by uncertainties in the cosmic $\psi$ \citep[e.g.,][]{Kobayashi_13}; there are significantly different scalings between the observed quantities and SFR with redshift predicted by \citet{Madau_14} and \citet{Madau_17}. Therefore, at least, to strengthen the validity of a varying $k_{\rm cc}$, more precise data within $z\approx1-2$ by future surveying is most certainly required. In particular, the Nancy Grace Roman Space Telescope has excellent prospects in pinning down the cosmic CCSN rate with multiply lensed images \citep{Petrushevska_16}. In parallel, theoretical approach must be upgraded to acquire a detailed composite of SFRs contributed from various galactic components as a function of $z$, which could assessed from the results of  cosmological numerical simulations reproducing the mix of early-type and late-type galaxies \citep[e.g.,][]{Vogelsberger_14}. In summary, there are large uncertainties in the argument on the IMF variability in terms of the cosmic CCSN rate as of now. Nonetheless, we place emphasis on the usefulness of our proposed framework for future studies. 

\section{Conclusions}

Recent findings that increase the likelihood that the upper mass bound for CCSN progenitor stars is as small as 18 \ms necessitate the close scrutiny of existing GCE models from this viewpoint. Accordingly, we find that the elemental abundance characteristics shaped by nearby thin disk stars strongly support this hypothesis, as do arguments for a top-heavy IMF in certain Galactic components$-$the bulge and the thick disk$-$formed by bursting star formation events in the early days of the Galaxy. This argument remains little changed even if we consider the theoretical framework that gives a complex explodability mapping as a function of stellar mass. Then, these findings yield new prospects for establishing a cosmic scaled relation between SFHs and supernova rates:~a high-$z$ Universe where stars were exclusively formed in early-type galaxies gives birth to a high frequency of CCSNe whose rate is higher than that estimated by a rise in the cosmic SFR from a low-$z$ Universe. This outcome, namely, a large contrast in the CCSN rates between $z\approx$0 and $z\approx$0.8, is indeed in good agreement with the measured rates. This proposed framework should be validated by future surveying that will precisely pin down the CCSN rate beyond $z\approx$1. One implication of this study is that a narrow CCSN progenitor mass range results in a high formation rate of BHs as a fate of failed supernovae (estimated to be $f_{\rm BH}\approx$29\%, 36\% at $z$$\approx$0 and 2, respectively). These high rates greatly influence the count of background diffuse supernova neutrinos, by enhancing their predicted flux \citep[e.g.,][]{Liebendorfer_04}, which would come within the capturing ranges of current and near-future detectors \citep{Abe_21}.

\section*{Acknowledgements}

The author gratefully acknowledge the anonymous referee for a careful reading of the manuscript and for providing constructive criticism that improved the work. The author also thanks M. Schultheis and A. Rojas-Arriagada for kindly providing the abundance data of the Galactic bulge. This work was supported by JSPS KAKENHI Grant Numbers 18H01258 and 19H05811.

\section*{Data Availability}

The data underlying this article will be shared on reasonable request to the corresponding author.









\bsp	
\label{lastpage}
\end{document}